\documentclass[%
 reprint,
 amsmath,amssymb,
 aps,
]{revtex4-2}

\usepackage{graphicx}
\usepackage{dcolumn}
\usepackage{bm}
\usepackage{color}
\usepackage{amsmath,amssymb}

\begin{document}

\title{Saffman-Taylor Fingers at Intermediate Noise}

\author{Dan Shafir}
\email{dansh5d@gmail.com}
\author{David A. Kessler}
 \email{kessler@dave.ph.biu.ac.il}
\affiliation{%
 Dept. of Physics, Bar-Ilan University, Ramat-Gan 52900 Israel
}%

\date{\today}

\begin{abstract}
We study Saffman-Taylor flow in the presence of intermediate noise numerically by using both a boundary-integral approach as well as the Kadanoff-Liang modified Diffusion-Limited Aggregation model that incorporates surface tension and reduced noise. For little to no noise, both models result reproduce the well-known Saffman-Taylor finger. We compare both models in the region of intermediate noise where we get occasional tip-splitting events, focusing on the ensemble-average. We show that as the noise in the system is increased, the mean behavior in both models approaches the $\cos^2(\pi y/W)$ transverse density profile far behind the leading front. We also investigate how the noise scales and affects both models.
\end{abstract}

\maketitle

\section{\label{sec:level1}Introduction}

Much research  has been dedicated to the problem of the Saffman-Taylor (ST) system \cite{SaffmanTaylor1958} which involves the displacement of a viscous fluid (water, say) by a relatively inviscid fluid (air) between two parallel plates in a effectively two-dimensional system first introduced by Hele-Shaw \cite{Hele-Shaw1898}.

As shown by Saffman and Taylor, an initially flat interface is intrinsically unstable and the interfacial surface tension plays the role of stabilizing the interface at short length scales. For not too small values of the dimensionless surface tension parameter, $d_0$, the resulting pattern is the famous Saffman-Taylor finger, with a width roughly one-half the channel width. For smaller surface tension and/or higher driving, the finger tip splits and then heals, due to the noise in the system. Under more extreme conditions, the splitting is so rapid that there are always multiple tips, and we get highly branched random structures, reminiscent of those generated by the Diffusion-Limited Aggregation (DLA) model~\cite{DLA1981}.

The governing equations for the motion of the interface have been formulated and solved numerically by various approaches, among them the Boundary-Integral method adopted herein-\cite{Advances1988, dendritic1986, BrowerKesslerKoplikLevine1984I, KesslerKoplikLevine1984II}. The model solves for the instantaneous local velocity of the interface instability via a Green’s function approach. The interface is then advanced in time numerically using the resulting velocity.

The second model we consider here is a variation on the DLA model first introduced by Witten and Sander~\cite{DLA1981}. In the classic DLA model, particles are released from far away and diffuse towards an existing aggregate, attaching to it on first contact. DLA generates intricate random branched structures. A variation by Kadanoff \cite{Kadanoff1985} and Liang \cite{Liang1986} on the classic DLA model, which we refer to as KL-DLA, incorporates the effects of surface tension and reduces the noise in the system. The resulting model at low noise levels very well reproduces the deterministic solutions of the Saffman-Taylor flow equations.

\begin{figure}
\centering
\includegraphics[width=1\linewidth]{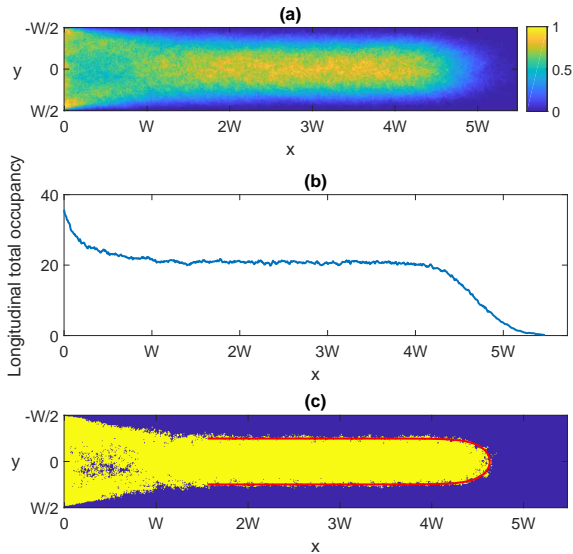}
\caption{An overview of the A. Arneodo and Y. Couder, et al.~\cite{DLA1989,Arneodo1991} results on classic DLA: (a) The occupancy density $r(x,y)$ with a channel width of $W=128$ of 500  DLA runs each with a total mass of 12800 particles. The color bar corresponds to the fraction of the runs in which a given cell was occupied by a particle. (b) Longitudinal total occupancy $r(x)=\int dy r(x,y)$ of pallet (a) as a function of $x$ (the growth direction). (c) Points where the occupancy density in (a) are larger then $1/2$. The solid line corresponds to the Saffman-Taylor solution for $\lambda=0.5$.}
\label{fig: DLA limit}
\end{figure}

An intriguing step toward characterizing the high-noise/zero surface tension DLA region was taken by A. Arneodo and Y. Couder, et al. \cite{DLA1989, Arneodo1991} who considered what they termed the occupancy density map, the fraction of runs in which a given cell is occupied by a particle, see Fig. \ref{fig: DLA limit}(a). They showed that for classic DLA, this map  is very well fitted by $\cos^2(\pi y/W)$, where $y$ is the axis perpendicular to the aggregate growth direction and $W$ is the channel width. This is true for a region of the occupancy density that has already stabilized far enough from its tip. Furthermore, when this occupancy density map is drawn only for cells visited in more than half of the runs, one obtains a shape whose outline is very well fitted by the Saffman-Taylor analytical solution \cite{SaffmanTaylor1958} for a finger width (compared to the channel’s width) of $\lambda=0.5$ as can be seen in Fig. \ref{fig: DLA limit}(c). Thus, classic DLA is seen to ``remember" its deterministic Saffman-Taylor origins, at least at the level of the occupancy density map.

The purpose of this paper is to use the two models discussed to track how, as we introduce stronger noise into both models, the occupancy density map approaches the limiting profile of the classic DLA. We also compare both models and investigate how similar they both are in the region of intermediate noise, region of noise where the shape of the interface is neither DLA-like nor does it resemble a stable finger. We also investigate how the noise is scaled in both models.

This article is structured as follows: In section \ref{section: basic equations} we review the basic equations defining the Saffman-Taylor model. In Sections \ref{section: The boundary integral method} and \ref{section: DLA} we present and discuss our results of the noisy boundary-integral method and the KL-DLA model respectively. In section \ref{section: Compare} we compare the two models, focusing on the noise dependence. In section \ref{section: conclusion} we present our conclusions.

\begin{figure*}
\includegraphics[width=\textwidth]{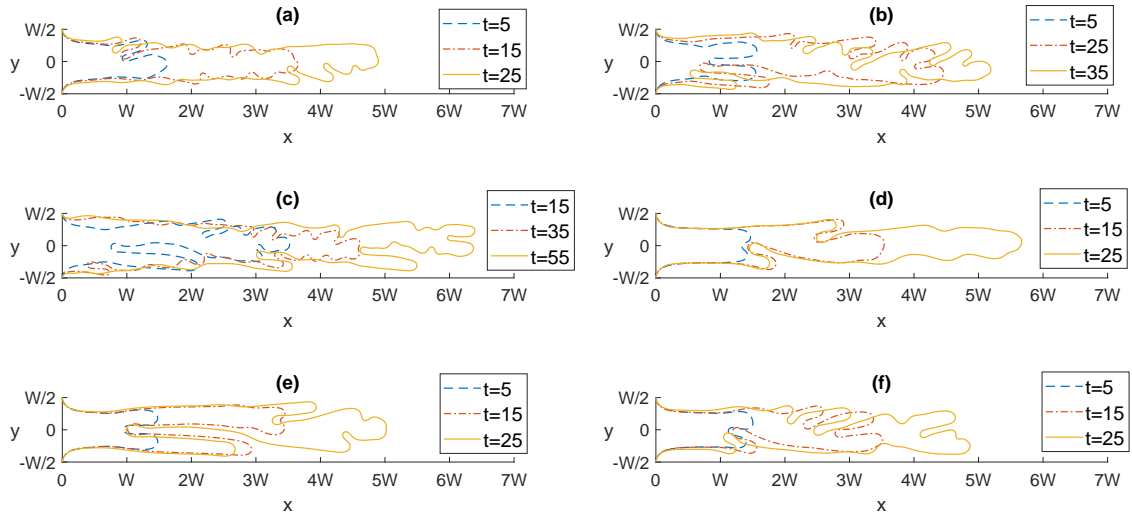}
\caption{\label{fig:finger examples}Boundary-Integral method finger progression with a channel width of $W= \pi$ for different values of noise magnitude parameter $f_0$ and the surface tension parameter $d_0$. The time step was taken to be $\Delta t= 0.05$. Runs calculated up to $t=25$ were done with $N=400$ equally spaced points in arc-length, up to $t=35$ with $N=600$ and up to $t=55$ with $N=800$. (a) $d_0=0.01,\:f_0=0.0625$. (b) $d_0=0.01,\:f_0=0.125$. (c) $d_0=0.01,\:f_0=0.25$. (d) $d_0=0.02,\:f_0=0.125$. (e) $d_0=0.02,\:f_0=0.25$. (f) $d_0=0.02,\:f_0=0.375$.}
\end{figure*}

\begin{figure*}
\centering
\includegraphics[width=\textwidth]{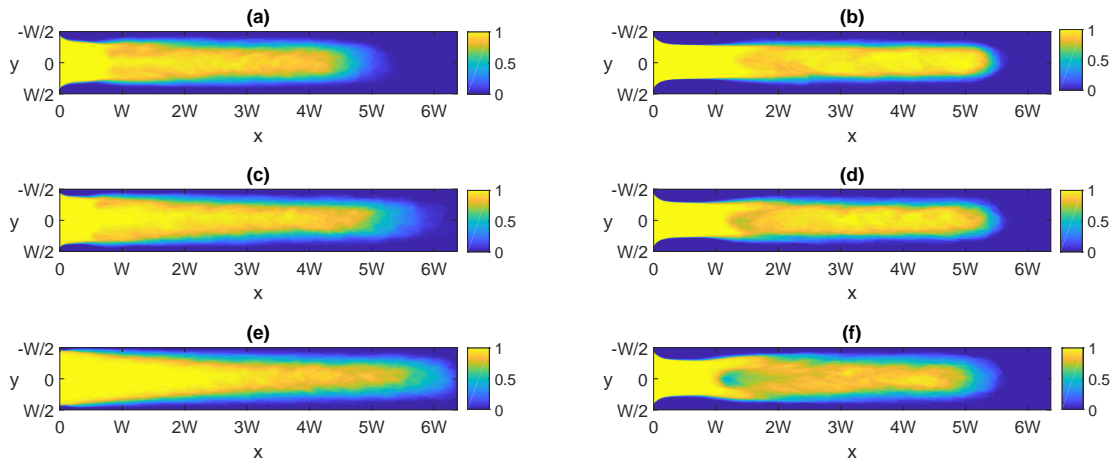}
\caption{Boundary-Integral method occupancy density maps of 120 independent runs with channel width $W= \pi$ for different values of noise $f_0$ and surface tension $d_0$. The plot is on a color-scale where $0/1$ signifies that in non/all of the runs the point is occupied. Time step was taken to be $\Delta t= 0.05$. Runs calculated up to $t=25$ were done with $N=400$ equally spaced points in arc-length, up to $t=35$ with $N=600$ and up to $t=55$ with $N=800$. (a) $d_0=0.01,\:f_0=0.0625,\:t=25$. (b) $d_0=0.02,\:f_0=0.0625,\:t=25$. (c) $d_0=0.01,\:f_0=0.125,\:t=35$. (d) $d_0=0.02,\:f_0=0.125,\:t=25$. (e) $d_0=0.01,\:f_0=0.25,\:t=55$. (f) $d_0=0.02,\:f_0=0.25,\:t=25$.}
\label{fig: finger occupancy density}
\end{figure*}

\begin{figure*}
\centering
\includegraphics[width=\textwidth]{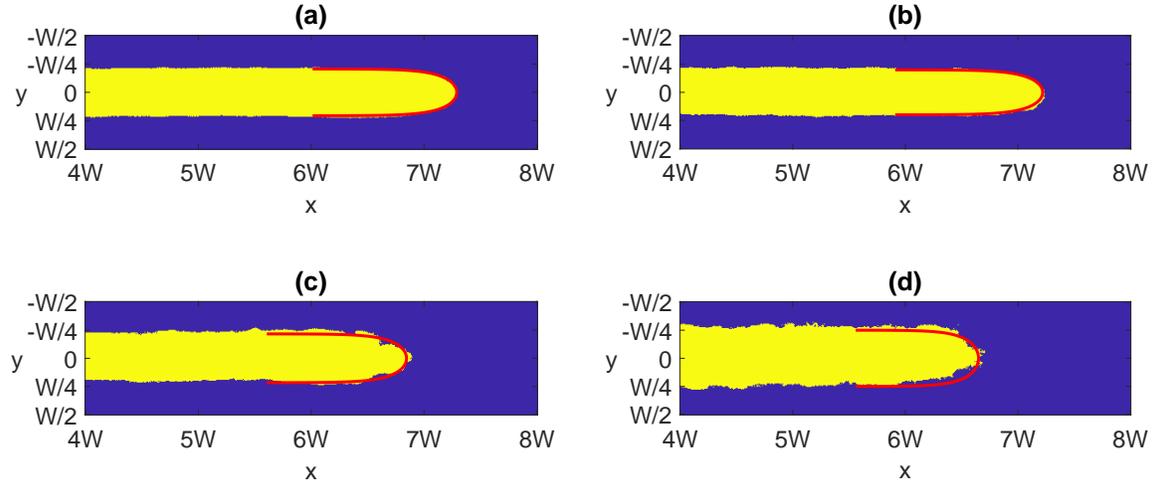}
\caption{Plots of the points where the occupancy density maps (normalized) in the KL-DLA case are larger then 1/2 for different values of the control parameter $B$ ($M=3$). Each plot consists of 120 independent runs of the same total mass. The channel width is $W=128$ cell units. Each graph's outline is fitted to the Saffman-Taylor analytical solution (the solid line) using the $\lambda$ value as a fitting parameter. (a) $B=0.002$ , $\lambda=0.415$. (b) $B=0.0015$ , $\lambda=0.4$. (c) $B=0.001$ , $\lambda=0.433$. (d) $B=0.0008$ , $\lambda=0.5$.}
\label{fig:KL-DLA outlines}
\end{figure*}

\begin{figure*}
\centering
\includegraphics[width=\textwidth]{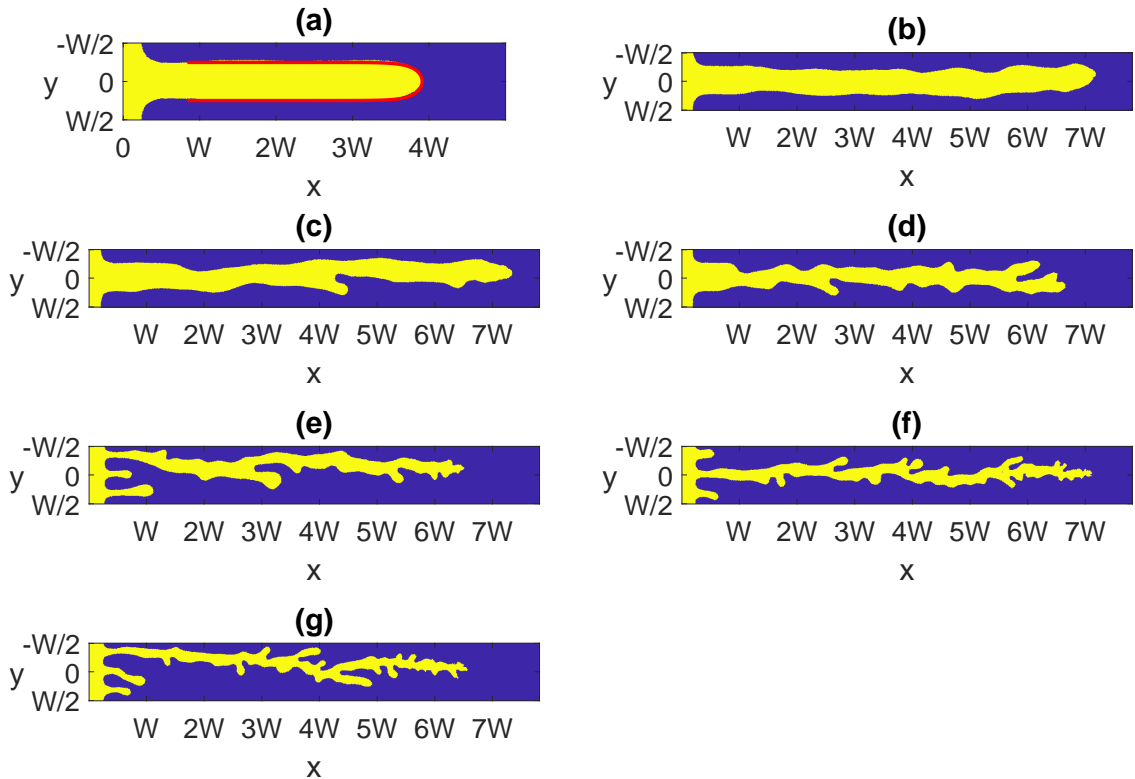}
\caption{KL-DLA outcomes for different values of $B$ and $M$. The channel width is $W= 128$ cell units. Runs (a)-(d) were initialized from the Saffman-Taylor analytical solution of $\lambda=1/2$. Runs (e)-(g) were initialized from a flat interface with a perturbation of wave length which equals to $1/3$ of the channels width. (a) $B=0.008,\:M=20$. The solid line corresponds to the saffman-Taylor analytical solution of $\lambda=1/2$. (b) $B=0.002,\:M=3$. (c) $B=0.0015,\:M=3$. (d) $B=0.001,\:M=3$. (e) $B=0.0008,\:M=3$. (f) $B=0.0006,\:M=3$. (g) $B=0.0005,\:M=3$.}
\label{fig: DLA examples}
\end{figure*}

\begin{figure*}
\centering
\includegraphics[width=\textwidth]{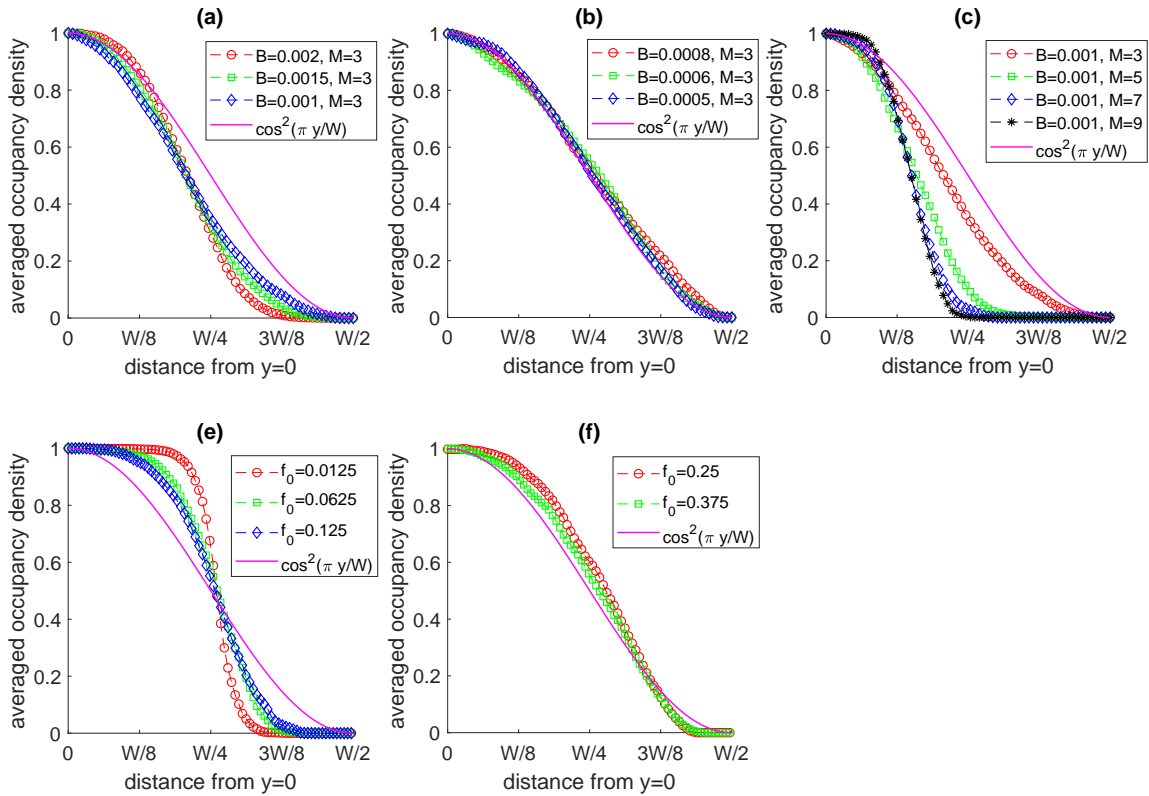}
\caption{Plots of the average occupancy density (normalized) for 120 independent runs compared to the limiting solution $\cos^2 (\pi y / W)$. Upper row: the KL-DLA case. To receive the average occupancy density maps we averaged on the interval $4W<x<6W$. (a), (b) are plots for changing $B$ with constant $M$. (c) are plots for constant $B$ and changing $M$. In all the DLA runs the channel width is $W=128$ and $L=11$. Bottom row: the Boundary-Integral method with surface tension $d_0=0.02$ for different values of noise level $f_0$. To receive the average occupancy density maps we averaged on the interval $10<x<14$. The channel width is $W=\pi$.}
\label{fig: avg occupancy density}
\end{figure*}

\begin{figure*}
\centering
\includegraphics[width=\textwidth]{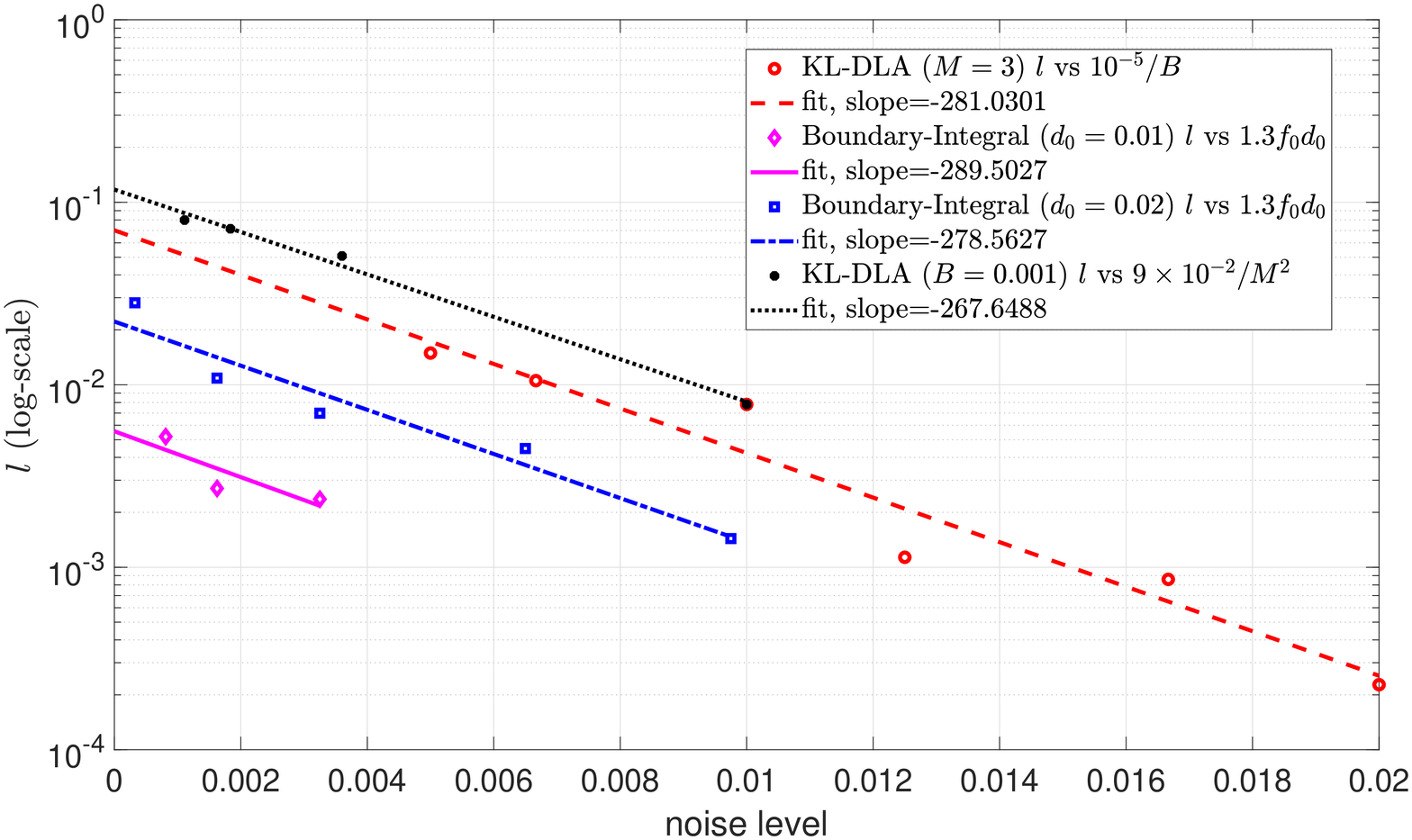}
\caption{ The quantity $l$, a measure of the distance to the Arneodo-Couder profile, for both models vs. the noise level, displayed in semi-logarithmic scale . The $l$ for the KL-DLA model is plotted for two cases, one with constant $M$ against $10^{-5}/B$, and the other with constant $B$ against $9 \times 10^{-2}/M^{2}$. The $l$ for the Boundary-Integral model is plotted against the noise magnitude variable $f_0$ times the surface tension variable $d_0$ (times $1.3$ in order to have the same approximate slope as in the KL-DLA plots). For each case we also plot its linear fit. Each point corresponds to an average on the occupancy density map of 120 runs. The point where $\text{noise level} = 0.01$ (and $M=3$) is shared by both the KL-DLA plots.}
\label{fig:loss compare}
\end{figure*}

\section{Basic Equations} \label{section: basic equations}
This section presents the governing equations of the Saffman-Taylor model~\cite{SaffmanTaylor1958} of Hele-Shaw flow. A Hele-Shaw cell \cite{Hele-Shaw1898} is a pair of glass plates arranged so that fluid flow takes place in a narrow gap of constant width between the plates. At first the gap is filled with water. Air is then pushed in to displace the water. The focus is on how the air-water interface develops. 
The governing equation of Hele-Shaw flow is Darcy's Law:
\begin{equation}
    \mathbf{v}=-\frac{b^{2}}{12 \mu} \mathbf{\nabla} p
\end{equation}
Here $\mathbf{v}$ is the velocity, $p$ is the pressure, $b$ is the gap thickness and $\mu$ is the viscosity.

If we consider incompressible viscous flow, and that for Darcy's Law the fluid velocity far downstream becomes asymptotically uniform in the $x$ direction (the downstream direction) we receive:

\begin{align}
           \nabla^{2} p &=0 \nonumber\\ 
           p_{x \rightarrow \infty} &\rightarrow-\frac{12 \mu}{b^{2}} v_{\infty} x
\end{align}

Considering the interface motion, the interface normal velocity is the normal component of the fluid velocity at the interface. Labeling the interface normal by   $\hat{n}$, we then have 
\begin{equation}
    -\hat{\mathbf{n}} \cdot \nabla p=\frac{12 \mu}{b^{2}} v_{\mathrm{n}}\,\,,
\end{equation}
where $v_n$ is the normal velocity of the interface. Given that the walls are impenetrable at $y=\pm \frac{W}{2}$ (W being the channel width), the boundary condition reads $\partial p / \partial y |_{y=\pm \frac{W}{2}} = 0$.

The pressure at the fluid boundary is given by the Young–Laplace equation:

\begin{align} \label{eq:pressure-boundary}
    p=p_{\mathrm{air}}-\gamma \kappa
\end{align}
where $p_{air}$ is the (constant) pressure of the inviscid fluid (in this case air), $\gamma$ is the surface tension and $\kappa(s)$ is the curvature of the interface at a certain point $s$. Shifting the pressure by $p_{\mathrm{air}}$ and converting to dimensionless variables by defining: 
\begin{align}
    \phi=-\frac{b^{2}}{12 \mu v_{\infty} (W/2)} \: p, \quad d_0=\frac{\gamma}{12 \mu v_{\infty}}\left(\frac{b}{W/2}\right)^{2}
\end{align}
and scaling all lengths by the half-channel width $(W/2)$, we get the final form of the Saffman-Taylor equations for the velocity's potential $\phi$: 
\begin{align} \label{Saffman-Taylor equations}
    \begin{aligned} \nabla^{2} \phi &=0 \\ \hat{\mathbf{n}} \cdot \nabla \phi &=v_{\mathrm{n}} \\ \phi\left(\mathbf{x}_{\mathrm{inside}}\right) &=d_0 \kappa \end{aligned}
\end{align}
with boundary conditions
\begin{align}
    &\phi \sim x \; as \;  x \rightarrow \infty \label{eq: boundary condition 1} \\ 
    &\frac{\partial \phi}{\partial y}\Biggr|_{y=\pm \frac{W}{2}} = 0
\end{align}
where $d_0$ is the dimensionless surface tension.

\section{The Boundary-Integral Method} \label{section: The boundary integral method}
In this section we present the Boundary-Integral method \cite{Advances1988, dendritic1986, BrowerKesslerKoplikLevine1984I, KesslerKoplikLevine1984II}, the way we introduce noise into the system and the results of comparing the occupancy density map to the limiting profile of the classic DLA.

\subsection{Equations}
In order to solve the Saffman-Taylor equations (eq. \ref{Saffman-Taylor equations}), we use an approach that converts the system to an integro-differential equation \cite{Advances1988, dendritic1986}:
\begin{align}\label{eq: integro-differential equation}
    \begin{split}
        \int G(x,x',y,y') v_{n}\left(s^{\prime}, t\right)&= \\ \int \hat{\mathbf{n}}^{\prime} \cdot \nabla^{\prime} &G(x,x',y,y') d_0 \kappa\left(s^{\prime}\right)+A
    \end{split}
\end{align}
Here $G(x,x',y,y')$ is the Green's function for the two-dimensional Laplace equation and $s$ and $s'$ are points on the interface perametrized by arclength, corresponding to locations $(x,y)$ and $(x',y')$ respectively. $A$ is a constant we will find later. This is an equation specifying the normal velocity $v_n$ given the position of the interface and its curvature $\kappa$.
The Green's function incorporates the no-flux sidewall boundary condition.
The Green's function can be obtained using the method of images with the result \cite{Advances1988, dendritic1986}:
\begin{widetext}
\begin{eqnarray}
 G(x,x',y,y')=-\frac{1}{2}(x-x')
 &&-\frac{W}{8\pi}\log
 \left[
   1-2\exp{\left(-\frac{\pi}{W}(x-x')\right)}\cos\left(\frac{\pi}{W}(y-y')\right)
   +\exp{\left(-\frac{2\pi}{W}(x-x')\right)}
 \right]\nonumber\\
 &&-\frac{W}{8\pi}\log
 \left[
   1+2\exp{\left(-\frac{\pi}{W}(x-x')\right)}\cos\left(\frac{\pi}{W}(y+y')\right)
   +\exp{\left(-\frac{2\pi}{W}(x-x')\right)}
 \right]\;.
\end{eqnarray}
\end{widetext}

In order to satisfy our boundary condition of constant velocity far down the channel (eq. \ref{eq: boundary condition 1}) we look for $v_n$ that satisfies:
\begin{align} \label{eq:Vcondition}
    \int v_{n}(s, t) d s=C
\end{align}
meaning we pump a constant amount of flux into the system at any given time to propagate the interface.  We choose units such that $C=1$.
Now we can find the constant $A$ in eq. \ref{eq: integro-differential equation} by looking for the value that satisfies condition \ref{eq:Vcondition}.

Next we parameterize the curve by $\theta(\alpha)$, the angle the normal vector makes to the flow direction as a function of relative arclength $\alpha=\frac{s}{S_T} , (0\leq\alpha\leq 1)$ and the total arclength $S_T$. Once we know the shape of the interface, we can find the normal velocity $v_n$ by realizing eq. \ref{eq: integro-differential equation} and condition \ref{eq:Vcondition}. We use $v_n$ to step the interface forward in time using the following equations \cite{BrowerKesslerKoplikLevine1984I}:
\begin{align}
    \begin{split}
    	\dot{\theta} (\alpha) &= \frac{1}{S_T} \frac{\partial v_n (\alpha)}{\partial \alpha} - \frac {\partial \theta}{\partial \alpha}[ \int_{0}^{\alpha} v _n \kappa d \alpha ' - \alpha \int_{0}^{1} v_n \kappa d \alpha ]\\
    	 \dot{S_T} &= S_T \int_{0}^{1} v_n \kappa d \alpha
    \end{split}
\end{align}
while the dot symbol is the derivative with respect to time. 

We verified that our code reproduces quantitatively the Saffman-Taylor instability of an almost flat interface and evolves into a stable Saffman-Taylor finger of the correct width.

\subsection{Adding Noise}
To help define the starting interface in our simulations we use the Saffman-Taylor analytical solution \cite{SaffmanTaylor1958}:
\begin{align} \label{eq:saffman-taylor finger-curve}
    x=\frac{1-\lambda}{\pi} \ln \frac{1}{2}\left(1+\cos \frac{\pi y}{\lambda}\right)
\end{align}
This equation describes the interface between the fluids as a function of $\lambda$ ($\lambda$ being the fraction of the channel occupied by the finger after the nose has passed).
We start each simulation from the upper part of the Saffman-Taylor analytical solution (eq. \ref{eq:saffman-taylor finger-curve})
with $\lambda$ taken to be $0.5$ connecting it to the walls of the channel using a quarter of a circle. Next we add noise to the system \cite{sidebranching1986}. We assume no symmetry and so evaluate the whole curve. After each calculation of the velocity (before time stepping the curve) we give the tip and its two adjacent points on the curve independent increments of the form:
\begin{equation}
    \begin{aligned}
        \delta v_n(i)/v_n(i)=f_0(2\mu-1)\\
         i=
         \begin{cases}
         tip \,\, index \\
         tip \,\, index+1 \\
         tip \,\, index-1
         \end{cases}
    \end{aligned}
\end{equation}
where $\mu$ is a random number uniformly distributed in the range (0,1), and $f_0$ is the noise amplitude. In Fig. \ref{fig:finger examples} we show a few typical results for different values of the noise parameter $f_0$ for $d_0=0.01$ and $d_0=0.02$. The fact that we add the noise always near the tip of the interface results in multiple side-branching of the finger. When a tip-splitting occurs, one side-branch of the two becomes more dominant over time, eventually developing into the main finger that continues propagating while the other, less dominant side branch, lags behind.

\subsection{Results and Discussion} \label{section: Boundary-integral results and discussion}
After many independent runs up to a certain time $t_f$ for a particular set of parameters (noise level $f_0$ and surface tension $d_0$) we want a quantitative way to represent the  outcome. For this we use the method described by A. Arneodo and Y. Couder, et al \cite{DLA1989}. We divide the space into a grid (in our case of width and height of $0.02$) and count for each cell the relative number of times it is occupied by air indicated by being inside the interface (the less viscous fluid injected into the cell) resulting in a grid of occupancy density $r(x,y)$, see Fig. \ref{fig: finger occupancy density}. We can see that as the noise level is increased the occupancy density becomes more smeared out.

These runs indicate, at least qualitatively, that as the noise level increases and the surface tension decreases in the case of the Boundary-Integral method (analogous to viscous fingering in a classical Hele-Shaw cell), the results become closer to those of the classic DLA model. We ran our simulation 120 times for increasing values of the noise strength $f_0$. For each set of 120 runs of a particular value of $f_0$ we computed the occupancy density map $r(x,y)$. Now for each occupancy density map we averaged over a section of $x$ where the pattern has stabilized resulting in a function we call the average transverse occupancy density $\bar{r}(y)$ which depends only on $y$. We can decide on a region of $x$ where the occupancy density map has stabilized by looking at the plot (as in Fig. \ref{fig: DLA limit}(b)) of the longitudinal total occupancy $r(x)=\int dy r(x,y)$.

In Fig. \ref{fig: avg occupancy density}, panels (e) and (f), we make use of the symmetrical characteristics of the problem and plot for each value of the noise parameter $f_0$ the average transverse occupancy density $\bar{r}(y)$ of the Boundary-Integral method vs. the distance from the line $y=0$. We can see  that as we increase the noise parameter $f_0$ with constant surface tension ($d_0$)  the average occupancy density $\bar{r}(y)$ approaches closer to the limiting solution $r=\cos^2 (\pi y / W)$ of A. Arneodo and Y. Couder, et al. \cite{DLA1989, Arneodo1991}. 

\section{DLA With Surface Tension and Reduced Noise} \label{section: DLA}
In this section we present the KL-DLA method proposed by Kadanoff~\cite{Kadanoff1985} and Liang~\cite{Liang1986} to introduce surface tension and reduced noise into the classic DLA model of Witten and Sander~\cite{DLA1981}. We discuss how the algorithm is constructed and we use the occupancy density map tool to show how, as the noise level in the system is increased, we approach the limiting solution $r=\cos^2 (\pi y / W)$ of A. Arneodo and Y. Couder, et al. \cite{DLA1989, Arneodo1991}.

\subsection{The Algorithm}
In the KL-DLA method there are two types of random walks involved in the simulation. Type one are particles that are being added into the system from “infinity” and represent the added flux from outside. Type two represent the surface tension and occur when a particle on the interface is involved in a rearrangement. The probability $p_r$ that a particle will detach itself and be involved in a rearrangement is given by~\cite{Liang1986}:
\begin{align} \label{eq: DLA pr}
    p_r(s)=\frac{T}{R_{s}}+p_{air}
\end{align}
where $R_s$ is the surface curvature at point $s$ and $T$ is the surface tension parameter. $p_{air}$ is the resting pressure of air which is irrelevant to the results and set to zero. The probability is always taken to be positive, the sign of $R_s$ determines if a particle detaches from the cell in question or is added to it. Thus, in a type two random walk, after choosing a cell that will be involved in a rearrangement (according to equation \ref{eq: DLA pr}) we release a particle and let it walk until it comes in contact with a cell on the interface. If the sign of $R_s$ is positive (negative) the particle will be removed (added) from the starting air-water boundary and be added (removed) to the ending boundary. In our simulation, we normalize the probability distribution of the curve so that the tip of the interface always has a probability of one to detach itself.

The relative frequency of these two kinds of walks is set by the dimensionless parameter $B$ which also determines $\lambda$, the ratio of the finger’s width to the width of the channel, $W$. This parameter plays the same role as the dimensionless parameter $B$ in Hele-Shaw flows defined by Trayggvason and Aref \cite{Tryggvason_Aref_1983} to be:
\begin{align}
    B=\frac{1}{W^{2}} \frac{T}{\frac{12 \mu}{b^{2}} v_{\infty}}=\frac{1}{W^{2}} \frac{T}{\nabla p_{x \rightarrow \infty}}
\end{align}
where reminding that $v_{\infty}$ is the velocity far down the channel, $p$ is the pressure, $b$ is the gap thickness and $\mu$ is the viscosity.

Following Liang~\cite{Liang1986}, we set $N_b$ to be the number of cells comprising the interface (equivalent to interface length). The probability of a walk of type one (walk starts at "infinity") will then be given by $1/(8 B N_b)$. Intuitively this makes sense since the longer the finger is, the more places there are that need rearranging. To reduce further the noise in the system and to be able to recreate a stable finger, we let a particle hit a cell multiple times before it is filled. Similarly, a particle needs to leave a cell multiple times before it is emptied. We set this variable to be $M$. Due to the noisy nature of the interface, the probability $p_r$ in Eq. \ref{eq: DLA pr} that a particle will be detached and involved in a rearrangement may be larger then one. This is considered as extra flux and is "carried away" (in addition to the base flux of one) by the particle to where it ends its random walk. This approach may cause the accumulated flux at the ending boundary site to be larger than $M$ (or smaller than $-M$). In that case, after moving the boundary we equally distribute the extra flux among the new neighbouring boundary sites. Since flux is always transferred from one boundary cell to another we ensure conservation of mass. 

The probability $p_r$ that a particle will detach and be involved in a rearrangement is proportional to the surface curvature at that point. To estimate the radius of curvature at a certain point $s$ we use a method first introduced by Vicsek \cite{Vicsek1984}. We count $N$, the number of unoccupied cells within a circle of diameter $L=2L_0+1$ centered at $s$. The linear connection between $N$ and the curvature of a discretized interface was analyzed and shown in various papers \cite{Bullard_1995, Frette_2009}. In order to set the zero curvature to the right value of $N$, we subtract the $N_0$ of a flat interface, which is equal to the total cells within a circle of diameter $L=2L_0+1$, divided by 2. Additionally, since we are looking at an occupied point \textbf{on} the interface, we get a bias from the discrete nature of the domain. This bias needs to be accounted for by weighting each cell on the interface by half. The value of $L$ is chosen in accordance to the curvature of the stable wavelength ($L=11$ in our calculations). 

The particles movement obey reflective boundary conditions. To save running time, particles that venture out too far from the finger’s interface are returned to the region of interest using pre-calculated probabilities as was proposed by Kadanoff \cite{Kadanoff1985}. To farther reduce the noise in the system, we use another approach by Kadanoff \cite{Kadanoff1985}. The "most immediate neighbourhood" of a site $s$ is defined to be the 8 adjacent cells. A particle that is added to the aggregate at a site $s$ is added to the boundary within the "most immediate neighbourhood" of $s$ which has the most amount of air around it. In the case of "equally good" sites we choose one at random. Similarly, a particle that is removed from the aggregate is removed from the site with the least amount of air around it.

\subsection{Tests of the Algorithm}
The parameter $B$, which signifies the competition between surface tension and noise, was checked against the linear instability analysis of Saffman and Taylor \cite{SaffmanTaylor1958} as was done by Liang \cite{Liang1986}. We also checked incompressibility and the effects of the surface tension by running a simulation that starts from an odd polygon shape and then let the interface rearrange itself until a circle was formed. Additionally, we recreated the stable finger case for $\lambda=0.5$ as can be seen in Fig. \ref{fig: DLA examples}(a).

\subsection{Results and Discussion}
From our simulations, it is apparent that the tip of the finger in the KL-DLA case seems to become unstable as the noise level is increased.
This is true of course of the Boundary-Integral method as well. Since deterministically the width of the finger never falls below one-half of the channel width, the surface tension rearrangement effect cannot keep up with the increasing flux to the tip.
Due to the amount of noise present in the system, sometimes a KL-DLA run resulted in an aggregate with “holes” in the region occupied by air. These artifacts were also noted by Liang \cite{Liang1986} and they were eliminated from our statistical analysis. 

When we consider the plot of the occupancy density map for cells visited in more than half of the runs (as in Fig. \ref{fig: DLA limit}(c)), there exists a threshold for $B (<0.001)$ for which the outline is well fitted by the Saffman-Taylor analytical solution of $\lambda=0.5$. For values of $B$ higher than the threshold but still smaller than the stable finger case ($B=0.008$), the outline resembles solutions with a lower value of $\lambda$. For an example of this occurrence see Fig. \ref{fig:KL-DLA outlines}.

Now we follow the same steps as in subsection \ref{section: Boundary-integral results and discussion} and for each set of values for $B$ and $M$ (in addition to those that can be seen in Fig. \ref{fig: DLA examples} ) we calculate the occupancy density map of 120 independent runs with the same mass. Now for each occupancy density map we average over a section of $x$ where the pattern has stabilized resulting in the average occupancy density function $\bar{r}(y)$ which depends only on $y$. In Fig. \ref{fig: avg occupancy density}, panels (a)-(c), we plot the resulting average occupancy density $\bar{r}(y)$ of the KL-DLA (for different values of $B$ and $M$) against the limiting solution $r=\cos^2 (\pi y / W)$ of A. Arneodo and Y. Couder, et al. \cite{DLA1989, Arneodo1991}. From Fig. \ref{fig: avg occupancy density} we can see that as we decrease $B$ and $M$ the average occupancy density approaches this limit.

\section{Comparing the two models} \label{section: Compare}
In this section we are interested in comparing the two models in the intermediate noise regime.  In particular, we focus on the approach to the limiting solution $\cos^2(\pi y/W)$ in the two models.

To evaluate explicitly how near we are to the limiting solution, we define the quantity $l$ to be the average squared distance of $\bar{r}(y)$ to the limiting profile:
\begin{align}
    l=\frac{1}{n} \sum_{i=1}^{n}\left(\bar{r}\left(y_{i}\right)-\cos ^{2}\left(\frac{\pi y_{i}}{W}\right)\right)^{2}
\end{align}
where $n$ is the resolution of the occupancy density map on the y axis (128 in the case of KL-DLA and 157 in the Boundary-Integral method).

We plot in Fig. \ref{fig:loss compare} this quantity $l$ in both models as a function of "noise". In the KL-DLA model, the noise is controlled by the parameters $B$ and $M$. In the Boundary-Integral method, the noise is controlled by $f_0$. We can see that in the KL-DLA case, the loss has a very similar exponential dependence on both $(1/B)$ and $(1/M^2)$. Decreasing the values $B$ and $M$ each adds more noise to the system and causes the model to more closely approach the limiting solution $r=\cos^2 (\pi y / W)$, reflected in a reduced value of $l$. Similarly, in the Boundary-Integral method, $l$ scales exponentially with the surface tension parameter $d_0$ and the noise amplitude $f_0$. This implies that the surface tension parameter controls how fast we converge to the limiting solution as the noise is increased in the system. Higher surface tension ($d_0$) results in faster convergence. At the same time it seems that decreasing $d_0$ with constant noise amplitude $f_0$ also gets us closer to the limiting solution.

\section{Conclusion} \label{section: conclusion}
We have presented two numerical models of the Saffman-Taylor instability in a channel, the variation on the classic DLA model by Kadanoff \cite{Kadanoff1985} and Liang \cite{Liang1986} and the Boundary-Integral method \cite{Advances1988, dendritic1986, BrowerKesslerKoplikLevine1984I, KesslerKoplikLevine1984II} with added noise. In both models with little to no noise, runs result in the stable Saffman-Taylor analytical solution \cite{SaffmanTaylor1958} for $\lambda \geq 0.5$. We showed (Fig. \ref{fig: avg occupancy density} and \ref{fig:loss compare}) that in both models, as we increase the noise in the system the average occupancy density map approaches the limiting solution $\cos^2(\pi y/W)$ of A. Arneodo and Y. Couder, et al. \cite{DLA1989, Arneodo1991}. Further more, the KL-DLA model has the same exponential dependence on both parameters $(1/B)$ and $(1/M^2)$ which control the amount of noise in the system. In the Boundary-Integral method, the approach to the Arneodo-Couder profile depends exponentially on the surface tension $d_0$ and noise magnitude $f_0$. In conclusion, the two different models exhibit the same quantitative behaviour and convergence rates towards a shared limiting solution. It would be interesting to investigate to what extend the regularized mean-field DLA model~\cite{MeanField} can capture this behavior.

\begin{acknowledgments}
This work was supported in part by the Israel Science Foundation, Grant No. 1898/17.
\end{acknowledgments}

\bibliography{mybib}


\end{document}